# A Comparative Analysis of MOOC - Australia's Position in the International Education Market

## Georg Peters
Department of Computer Science & Mathematics
Munich University of Applied Sciences
Munich, Germany;
Australian Catholic University
Sydney, Australia
Email: georg.peters@hm.edu

## Doreen Sacker
Department of Computer Science & Mathematics
Munich University of Applied Sciences
Munich, Germany
Email:doreen.sacker@hm.edu

## Jan Seruga
Faculty of Education and Arts
Australian Catholic University
Sydney, Australia
Email: jan.seruga@acu.edu.au

## Abstract

Tertiary education is one of the most important industries in Australia and a crucial source of income for its universities. Therefore, any new form of tertiary education that may have impact on Australian universities should be closely watched for implications on their business models. In the past decade, MOOC-Massive Open Online Courses have emerged with the promise to offer world class education accessible anywhere and anytime for free or at very reasonable costs. MOOC not only provide opportunities for lifelong learning, they are also designed to enrich, compliment and even (partly) substitute classic tertiary education. Hence, it is of crucial importance how and where Australia takes its stand in this rapidly evolving market. In this paper, Australia's international position as a provider of MOOC is analysed and compared, in particular to Germany and the US.

## 1   Introduction

As observed in virtually any industry, information technology has also impacted the educational sector in the past decades. Although IT support has already led to a significant transformation, the process of change has been steady rather than an abrupt revolution. The distribution of static teaching material has changed from paper to PDF, teaching material has been enriched by interactive elements and recently social media has impacted the communication between teachers and students as well as between students themselves. However, a few years ago, a new business model in education was introduced: MOOC-Massive Open Online Courses. The fundamental idea behind MOOC is to offer high quality education to everyone via the web, i.e., free of charge and independent of time and geographical location. If this concept succeeds it will dramatically change how education is provided, in particular for experienced learners in the tertiary educational sector and for professional lifelong learning. Australia is one of the leading providers of tertiary education with several universities of excellent international reputation attracting students from all over the world. With their tuition fees these students heavily contribute to the income of the universities. Furthermore, international students have significant direct and indirect positive effects on the Australian economy as its whole, e.g., by spending money in Australia or by luring their relatives as tourists to Australia (Group of Eight Australia 2014). Last but not least, they are a source of a highly qualified workforce for Australian companies and function as a link between Australia and their home country.

MOOC could have the potential to significantly change international student mobility when high-quality education is accessible online anytime anywhere at considerably lower costs. Hence, it is crucial for the Australian tertiary education sector to understand and carefully watch the MOOC business models and potentially develop strategies to adapt to possibly new emerging market structures.

In the paper, we concentrate on the supply-side of the MOOC market and analyse its content providers and platforms. The remainder of this article is organized as follows. In the next section, we present a brief overview of Massive Open Online Courses. In Section 3, we provide a data summary and discuss preliminary findings. Then we conduct a comparative analysis of MOOC designed by American, Australian, and German universities and develop a MOOC Leadership Matrix. The paper concludes with a discussion and summary in Section 5.

## 2  Foundations of Massive Open Online Courses

### 2.1  Typology of MOOC

#### 2.1.1  Characteristics of Massive Open Online Courses

Siemens and Downes are regarded as pioneers offering MOOC as early as 2008 (Parry 2010). The characteristics of Massive Open Online Courses can be directly derived from the term itself. For our brief discussion we just change the order of words: (1) course, (2) open, (3) online, and (4) massive.

(ad 1) MOOC are courses, i.e., they have defined start and end dates. This ensures that students can form classes so that they can join learning groups to discuss the course material, support each other etc. The courses are designed by academics as well as by professionals from industry. (Hoy 2014)

(ad 2) MOOC are open with respect to several dimensions (Stewart 2013). Generally, the courses are offered free of charge (Hoy 2014). However, there might be fees when students strive for certificates or credits issued by a MOOC provider (Rodriguez 2013). Anyone can register independently of her/his qualification, and even when prerequisite knowledge is recommended it is not formally checked (Hoy 2014).

(ad 3) Courses are accessible for students only online without any face-to-face phases. The communication between the students is fostered by modern information technology applications, in particular by social media. These applications support the students to develop supplementing material themselves and share their ideas, challenges and learning documents with their classmates. (Jacoby 2014)

(ad 4) The characteristics of MOOC, open and online, make it possible that a practically unlimited number of students can take part in a MOOC. The word massive represents the opportunity that virtually everybody can join a course. For example, in 2011, more than 160,000 students enrolled for Thrun's MOOC "Introduction to Artificial Intelligence" (Jacoby 2014). Furthermore, the term also refers to the massive amount of knowledge that is generated by the students (Stewart 2013).

#### 2.1.2  Types of MOOC and Their Stakeholders

Basically, there are two main types of MOOC, namely cMOOC and xMOOC (Hoy 2014, Jacoby 2014, Rodriguez 2013, Stewart 2013): in the term cMOOC the c stands for "connective" and refers to the core characteristic of this teaching approach: connectivity. cMOOC foster learning as a process where individuals or groups autonomously form (temporary) communities or clusters. The objective of these communities is to learn in a group by distributing content and information, by mutually supporting each other and by sharing knowledge and ideas. In contrast to this, xMOOC are based on a more traditional idea of teaching and learning that is enriched by instruments of modern information technology. So, the letter x originates from the attribute "extended". A course is characterised by a pre-defined content and supporting course material which is very much in line with the still dominating behaviourist approach to teaching.

For our analysis, we identify three stakeholders in a narrow sense; platform providers, content providers and students. Furthermore, we fuzzily define society as stakeholders in a boarder sense. A challenge for MOOC platform and content providers is how they define their business model as long as the courses are open, particularly in a sense that they are free of any charge. Due to very high number of students, classes can be considered as highly heterogeneous, ranging from students without any previous access to high class education before (restricted by their location, qualification or funding), university students supplementing their knowledge to professionals looking for convenient possibilities for high quality lifelong learning. The fuzzily defined term society subsumes, e.g., countries and companies that rely on highly qualified workforces to remain competitive in a global economy. In the next sections, we focus on supply side, the MOOC platform and content providers.

## 2.2 MOOC Providers

In the past few years, MOOC platforms have been launched all over the world. The US is undoubtedly the leading country so far with platforms like Coursea, edX or Udacity. We enrich our brief survey on MOOC platforms by Australia and two European countries, Germany and the UK.

The success of Thrun and Norvig's MOOC "Artificial Intelligence" motivated Thrun to launch the for-profit company Udacity as a MOOC platform in 2011 (Rodriguez 2013). The courses on Udacity focus on classic STEM fields (science, technology, engineering and mathematics). In particular, Udacity addresses lifelong learning for professionals rather than the tertiary educational market. Hence, most of the 39 courses offered on Udacity are provided by technology companies like Google, Autodesk or NVIDIA. In total the platform has eleven partners currently (Udacity 2014).

In 2012, Harvard University and MIT founded edX as a non-profit open source platform (edX 2014). Like for any other platform, the main purpose of edX is to provide a system that hosts MOOC. Furthermore, Harvard University and MIT explicitly use edX as a research project to understand if and how IT may change the process of learning (edX 2014, Rodriguez 2013). Presently, selected courses at edX offer verified certificates or so called honour code certificates on the successful completion of a course (edX 2014, Rodriguez 2013). Currently, 53 partners offer more than 200 courses on edX in areas like humanities, mathematics, computer science and physics (edX 2014). In 2012, two Stanford University computer scientists, Koller and Ng, founded Coursera as a for-profit start-up (Coursera 2014). Coursera provides a MOOC platform for universities to develop and run their MOOC. Its community consists of around 100 partnerships, including the University of Melbourne, the University of Western Australia, Munich's Technical University and the Ludwig-Maximilians University. More than 22 million students from 190 countries are registered for courses at Coursera (Coursera 2014).

In 2012, the Open University in the UK launched FutureLearn as a for-profit company (Freitas 2013). Currently, 8 courses are in progress and another 45 courses announced. In some courses students can take tests to obtain certificates. FutureLearn partners with 40 universities and institutions, e.g., Monash University (FutureLearn 2014). An example for a MOOC platform from Germany is iversity. According to iversity (2014), it is first MOOC platform, where students can get credits within the ECTS framework (European Credit Transfer and Accumulation System). Another example is the German platform OpenCourseWorld founded in 2012. The first courses were offered in cooperation with universities in Saarbruecken, Munich and Hamburg. Generally, students can participate in the courses for free; however, for selected courses, certificates are offered for a fee (OpenCourseWorld 2014).

Open Universities Australia founded the MOOC platform Open2Study to give universities the opportunity to offer MOOC for free (Open Universities Australia 2014). Presently, more than a quarter of a million students participate in 49 courses in diverse areas like arts, humanities and education, science and engineering, and business. The courses are offered by 19 content providers including universities and companies. They subsume university courses, diplomas and certificates, free courses and professional training. (Open2Study 2014/2015) Another Australian platform, OpenLearning, was founded by Brimo, Collien and Buckland from the University of New South Wales in 2012. In 2014, it had more 40,000 students registered in 88 courses (OpenLearning 2014). In the meantime, it has grown to more than 200,000 students taking part in almost 450 courses. (OpenLearning 2015)

# 3 Data Summary

## 3.1 Methodology and Research Questions

Most of the platforms have been launched in the past five years. In our comparison, some leading platforms, Coursera, edX, FutureLearn, and Udacity, are analysed to give a first impression on MOOC. Note, due to the dynamically evolving market the provided figures can only be snapshots that may change fast. The data were collected from publically available sources, primarily from the webpages of the MOOC providers in late 2014.

The variety of courses offered by the platforms is an indicator for their diversity. We assume that the number of partners is an indicator for the growth potentials of a platform, particularly for the number of courses that may be offered in future. We summarize if the business models are for-profit or non-profit. It would be interesting to compare the number of students participating on each platform's website. However, the platforms do not publish sufficient information to analyse the number of students in-depth.

### 3.2 Findings

In order to compare the size of the platforms, the numbers of courses on the platforms have been analysed. In Table 1, some important figures of the MOOC are summarized.

| Platform | Based in | Estd. | Business Model | Num. of Courses | Num. of Partners | Courses / Partners |
|---|---|---|---|---|---|---|
| Coursera | US | 2012 | for-profit | 571 | 100 | 5.7 |
| edX | US | 2012 | non-profit | 200 | 53 | 3.8 |
| FutureLearn | UK | 2012 | for-profit | 53 | 40 | 1.3 |
| Udacity | US | 2011 | for-profit | 39 | 11 | 3.5 |

*Table 1: Key Figures of Selected MOOC Platforms*

A wider range of courses provides the opportunity to cover a wider range of topics and also attracts more participants. Coursera offers the largest number of courses in this comparison, with 571 courses compared to edX with just around 200 courses. FutureLearn offers only 53 courses and Udacity just 39.

Furthermore, the number of partnerships is an indicator for the growth potential of a platform. More partners are more likely to offer more courses. The comparison shows that Coursera, which has the most partnerships, already has the largest number of courses. Coursera has nearly twice as many partnerships with universities than the platform edX. Table 1 shows the relation between the number of courses and the number of partnerships. It can be observed that Coursera has the most active partners with each partner contributing 5.7 courses on average.

The comparison also shows that the majority of the platforms are based in the US. FutureLearn is an example for a platform from the UK. All platforms were launched around the same time in 2011 or 2012. The majority of platforms (Coursera, FutureLearn and Udacity) were founded as for-profit companies. In contrast to this, Harvard University and MIT founded edX as a non-profit organisation. As mentioned before, Harvard University and MIT offer courses on edX, but the platform is also intended as a research project that addresses the impact of information technology on learning.

Generally, MOOC platforms do not have a uniform model how to finance their businesses. However, some offer courses for free but charge for premium services and certificates. Like for several internet businesses, e.g., Twitter, we consider financing as one of the core challenges for MOOC providers.

In summary, Table 1 shows that Coursera is the leading MOOC platform currently. It hosts the largest number of MOOC and has the highest number of partnerships as well as the most active partners. Coursera is followed by edX. FutureLearn and Udacity are by far smaller platforms.

## 4 Analysis of MOOC offered by Universities

### 4.1 Methodology and Hypotheses

The following comparison between American, Australian, and German universities gives an exemplified snapshot for the development of MOOC. More and more universities are engaging in their development. In order to analyse the progress of MOOC in America, Australia and Germany, we selected the 10 top ranked universities of each country according to ARWU - Academic Ranking of World Universities (Shanghai Ranking Consultancy 2014). These universities have outstanding international visibilities, and leveraging on their reputation they probably play leading roles in the MOOC market. To compare them, data has been collected and analysed with respect to the number of MOOC the universities offer, when they established their first MOOC and the platforms they cooperate with. In addition, it has been analysed if the universities award students credits for completing a MOOC.

MOOC are still in an early stage of development. The first MOOC which is taken into account in this comparison is Thrun and Norvig's "Artificial Intelligence" at Stanford University in 2011. In Australia, the University of New South Wales set-up the first MOOC in 2012. In 2013, Ludwig-Maximilians University launched the first MOOC in Germany (Table 2).

The number of MOOC provided by the top ranked American, Australian and German universities have been collected to give an overview on the distribution of MOOC. The analysis shows that US universities offer by far the highest number of courses. These 10 American universities offer a total of 181 MOOC. The Australian universities offer 40 courses and the German universities only offer 13 MOOC.

| Country | Leading Platforms | | First MOOC | Growth Indicator | Number of MOOC | Unis with Credits |
|---|---|---|---|---|---|---|
| US | Coursera | edX | 2011 | 90% | 181 | 0 |
| AU | Coursera | edX | 2012 | 80% | 40 | 0 |
| GE | Coursera | iversity | 2013 | 40% | 13 | 3 |

*Table 2: MOOC in Selected Countries*

Furthermore, we discuss the number of universities that are active in the MOOC business presently. We think that more universities have the capacity to offer more courses - even they are choosing not to currently. Our analysis already showed that platforms with more partnerships offer more courses. The percentage of the 10 American, Australian, and German universities offering MOOC functions as an indicator for the possible growth of MOOC in each country: universities that already offer MOOC may leverage on their experience and launch further courses in future. Consequently, we define the percentage of universities currently offering MOOC as Growth Indicator.

Most of the universities do not offer credits for MOOC so far. Only in Germany can students obtain credits for completing a MOOC at the majority of the universities (three out of four universities). At the Technical University Munich students get credits for a MOOC in English if they do additional homework and a test as well. The University of Hamburg and the University of Kiel offer their students credits for the ECTS system through their MOOC. American and Australian universities do not offer credits for MOOC yet (Table 2). However, many of the universities offer verified certificates, such as the Australian National University, the University of Queensland, and the University of New South Wales.

| Platform | Coursera | edX | Open 2 Study | iversity | Future Learn | Class 2 go | Open Learn. | Stanford |
|---|---|---|---|---|---|---|---|---|
| Num. | 10 | 9 | 2 | 2 | 1 | 1 | 1 | 1 |

*Table 3: Number of Universities Offering MOOC on Platforms*

Table 3 shows that Coursera is the leading platform with respect to its network. It has the largest number of partners: 10, i.e., a third, of the universities offer their MOOC on Coursera. edX follows with 9 universities offering courses on its platform. The other platforms are far behind and partner with significantly fewer universities (note, Udacity, which was compared earlier with Coursera, edX and FutureLearn, has been left out in this analysis, because the partners on Udacity are mainly IT companies). In America as well as in Australia, Coursera and edX have the same number of universities as partners. In Germany, the number of partnerships for Coursera and iversity is identical. Hence, Coursera is the (jointly) leading platform in all 3 countries.

The following hypotheses are developed to highlight the main research findings (note, due to the small numbers we cannot statistically test the hypotheses but only indicate directions):

Hypothesis 1 – World Rank & MOOC. As mentioned above, we choose the top 10 universities of each country. On average, US universities are ranked significantly higher than German and Australian universities. Our data shows that American universities offer most of the MOOC. It is expected that higher ranked universities offer more MOOC than lower ranked universities:

- H1a. Higher ranked universities offer more MOOC than lower ranked universities.
- H1b. Lower ranked universities offer less MOOC than higher ranked universities.

Hypothesis 2 – Experience. American universities were the first to produce MOOC. Also 90% of the top American universities offer MOOC, compared to 80% of the top Australian, and 40% of the top German universities. In conclusion, it is believed that American universities offer more MOOC and are more experienced than Australian and German universities:

- H2a. On average American universities offer more MOOC than their Australian and German counterparts.

Hypothesis 3 – MOOC & Population & Students. The data discloses that American universities offer more MOOC than Australian and German universities. Furthermore, Australian universities offer more than double the number of MOOC in comparison to German universities. We expect that American universities offer more MOOC per citizen and student than Australian and German universities. Also, Australian universities are believed to offer more MOOC per citizen and student than German universities.

- H3a. American universities offer more MOOC per citizen than Australian or German.
- H3b. Australian universities offer more MOOC per citizen than German universities.
- H3c. American universities offer more MOOC per student than Australian or German.
- H3d. Australian universities offer more MOOC per student than German universities.

## 4.2　Findings

Hypothesis 1 – World Rank & MOOC. To test the first hypothesis the relation between the ARWU World Rank of the universities and the number of MOOC produced by the universities is analysed. We counted the number of MOOC of each university and related this number to the world rank of the corresponding university (Figure 1). The data shows that American universities are ranked significantly higher and offer more MOOC than Australian and German universities. In order to get a descriptive presentation in Figure 1, the American universities are depicted separately (right subfigure) from their Australian and German counterparts (left subfigure). Please note, the different scaling in the subfigures.

H1a. The majority (6 out of 10) of the German universities do not offer MOOC currently. 80% of the Australian universities already offer MOOC. Their positions in the world ranking vary widely from the University of Melbourne at #44 to Curtin University ranked between 300-400. For the Australian and German universities, the figure shows that there is no obvious correlation between the ranking and the number of MOOC. In Australia, the University of Sydney does not offer MOOC despite being highly ranked. In Germany, the positions in the world ranking of the universities that offer MOOC also show little correlation. However, for American universities there seems to be a link between their world rank and the number of MOOC they are offering: higher ranked American universities often offer more MOOC than lower ranked universities.

H1b. The results observed for high ranked universities are consistent with the results observed for low ranked universities. Low ranked often offer less MOOC than high ranked US universities. Note, low ranked is relative: low ranked within a country's top ten universities.

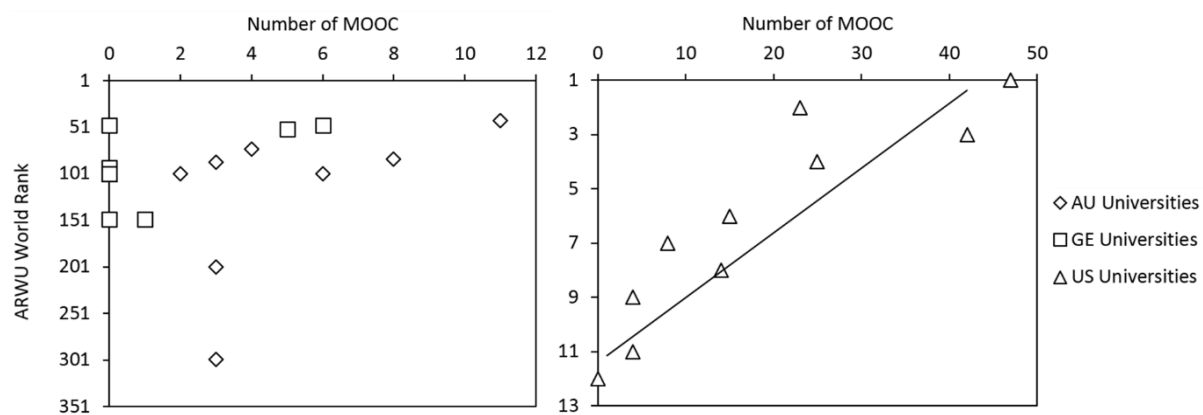

*Figure 1: Number of MOOC and World Ranking*

Hypothesis 2 – Experience. The number of MOOC offered by each university is collected, as well as the number of universities which offer MOOC in each country. In America 9 out of 10 universities offer MOOC, in Australia 8 out of 10 and in Germany 4 out of 10. To show how experienced an average university in each country is the total number of MOOC in each country is divided by the number of universities offering MOOC in each country. Table 4 shows that an American university on average has a portfolio of 22.22 MOOC. In contrast, an Australian university offers only 5.0 MOOC on average and a German university only 3.25. The hypothesis can be confirmed - American universities offer more courses on average compared to Australian and German universities.

| Country | Average Number of Courses | Population | Students | MOOC/ Citizen | MOOC/ Student |
|---|---|---|---|---|---|
| US | 22.2 | 317,260,000 | 24,800,000 | 5.7 E-07 | 7.3 E-06 |
| AU | 5.0 | 23,425,700 | 1,313,776 | 17.1 E-07 | 30.4 E-06 |
| GE | 3.3 | 80,800,000 | 2,499,409 | 1.6 E-07 | 5.2 E-06 |

*Table 4: Comparison of MOOC in Selected Countries*

Hypothesis 3 – MOOC & Population & Students. To give insights on relations between the numbers of MOOC and the size of America, Australia and Germany, the number of citizens and students in each country is considered. Hence, the total number of analysed MOOC in each country is divided by the number of citizens and students in each country.

H3a. Table 4 shows the population of each country (Statista 2014; Australian Bureau of Statistics 2014; Statistisches Bundesamt 2014). America has the biggest population, followed by Germany and Australia. As depicted in Figure 1, Australian universities offer the largest number of MOOC per citizen followed by the US and Germany. Thus, our hypothesis cannot be confirmed, as Australia offers more than twice as many courses per citizen than America and about ten times as many as Germany.

H3b. Australian universities offer more MOOC per citizen than both American and German universities. In conclusion, one can confirm this hypothesis.

H3c. To analyse this hypothesis, the number of students in 2013 was collected (Statista 2014; Australian Government Department of Education 2014). Table 4 shows that the gap between American and Australian universities is even bigger in comparison to the numbers of citizens. American universities offer less MOOC per student than Australian universities.

H3d. Australian universities offer significantly more MOOC per student than German universities (see Table 4).

### 4.3 MOOC Leadership Matrix

The findings so far show that American universities are the leading content providers significantly outpacing their Australian and German counterparts in absolute numbers. However, Australia is leading when the numbers are put in relation to a country's size.

To present the results in a condensed and illustrative way, we propose the MOOC Leadership Matrix by adapting the idea of the quadrants of the famous Boston Consulting Matrix (Henderson 1970).

| | **Exclusive Universities.** Leading, highly ranked universities, but considered reserved in the topic of MOOC because of the small number of MOOC. | **MOOC Leaders.** The universities are progressive in the topic of MOOC. They are leading in education worldwide and engage in online learning. |
|---|---|---|
| ARWU Rank | **Reserved Universities.** Universities which are not highly ranked and are reserved in the topic of MOOC. For example smaller universities with a lower budget. | **Advanced Universities.** Lower ranked universities that show a strong engagement in the topic of MOOC. |

Number of MOOC

*Figure 2: Definition of the MOOC Leadership Matrix*

Obviously, quality and reputation of a university are crucial factors when choosing a course. Hence, the first dimension of the MOOC Leadership Matrix consists of the ARWU World Rank of a university. The second dimension shows how experienced a university already is in offering MOOC. So, this dimension comprises of the number of MOOC a university is offering presently. In Figure 2, the MOOC Leadership Matrix is depicted and quadrants are defined.

When we apply the MOOC Leadership Matrix to we get results as shown in Figure 3 (note, as mentioned before, American universities are significantly higher ranked than Australian and German universities; therefore, American universities are shown in the right subfigure and Australian and German universities are presented separately in the left subfigure). Figure 3 impressively shows the leading positions of the US universities with respect to both dimensions, world ranking and number of MOOC. In a comparison of America, Australia and Germany virtually all of them are categorised as MOOC Leaders (note the different scaling in the subfigures). When we analyse the US market separately we observe that the two universities classified as MOOC Leaders are Harvard University and MIT. Both founded the platform edX and are pioneers in online learning. Most of the American universities are Reserved Universities, i.e., small numbers of MOOC and the lower world ranks. However, we need to emphasise again that this is inner US analysis; in terms of their international position, virtually all ten US universities are MOOC Leaders.

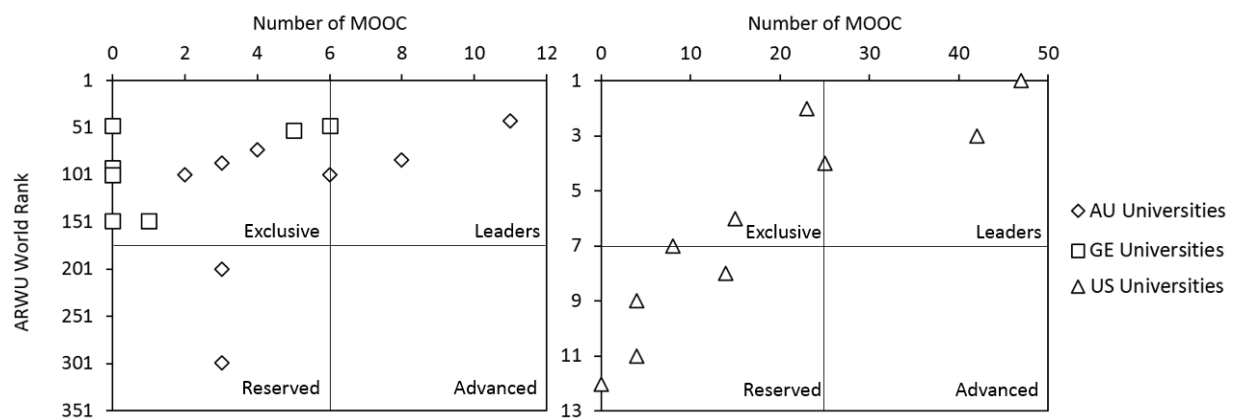

*Figure 3: MOOC Leadership Matrix: Comparative Analysis*

When we compare Australian and German universities, two Australian universities are classified as MOOC Leaders - the University of Melbourne, and the University of Queensland. Nearly all German universities are classified as Exclusive Universities. The 10 selected German universities are on average ranked higher than the Australian universities: 99.8 (Germany) and 124.5 (Australia). Hence, two Australian universities are considered as Reserved Universities. Last but not least, no lower ranked university offers an above average number of MOOC which would imply a classification as Advanced University.

## 5   Discussion and Conclusion

The concept of MOOC and the first courses as well as the first platforms were introduced in the US. To date, platforms and content providers from the US play a dominate role in the MOOC market. Coursera is also market leader in Australia (jointly with edX) and in Germany (jointly with iversity). US universities are the leading content providers. Most notable, Harvard and MIT offer more than 40 courses each on edX (edX 2014). Regarding demand, the courses offered by US universities are also far ahead. The enrolments of the most popular courses go into hundreds of thousands, e.g., Harvard's "Introduction to Computer Science" with around 285,000 students (edX 2014) or a class in social psychology on Coursera with 259,969 students (Walters 2014). Some important reasons for the dominating role of the US are obvious. The US is the leading country in the internet sector. So, it is not surprising that the idea of MOOC was developed and rapidly adapted by universities and start-ups in the US. Regarding demand, for the first time in history, unrestricted access to world-class education is possible. In terms of MOOC, the entry barriers to the world's leading universities - very competitive acceptance rates, tuition fees, intellectual abilities, geographical location - do not exist anymore. Why should students take second best when they can strive for the very best? Their desire might be fuelled by curiosity to see how a world-class university teaches; or they are driven by the vanity - or perhaps better, by the illusion – to become a member of one of the elite universities. Furthermore, the MOOC business is characterized by economies of scale and network effects (Stewart 2013): Regarding economies of scale, setting up and running a MOOC platform has quite high initial and fixed running costs but rather low variable costs. The same applies to the design and administration of a course. Beneficial network effects include examples such as, the larger the number of students in a course the greater the positive effects from social media. So, the US platform and content providers benefit from both, economies of scale and network effects.

Regarding MOOC platforms, iversity is the preeminent German provider. However, it is of negligible international visibility and importance. The leading universities are Ludwig-Maximilians University Munich and Technical University Munich. In terms of courses, German universities offer 13 MOOC only. However, German universities are leading with respect to credits: even though they have the smallest number of MOOC the majority of the courses allow students to get credits in contrast to their American and Australian counterparts. This may indicate that the German universities are pursuing a different strategy. While US providers position their MOOC as separated from their traditional educational business, German universities may understand them as complementing their classic teaching portfolio. A limitation for Germany's international position in the MOOC market might also be German as the regular language of instruction.

Currently, in Australia, the University of Melbourne is leading, offering 11 MOOC on Coursera, followed by the University of Queensland with 8 MOOC on edX. Eight out of 10 Australian universities offer MOOC (except the University of Sydney and the University of Adelaide which announced MOOC for 2015). So, where does Australia stand in the international MOOC market presently? Regarding its size, Australia is in an excellent position outpacing the US and Germany in both numbers of MOOC/citizen and MOOC/student. This might be an indicator that Australian universities traditionally need to be very internationally minded and attract students from abroad. However, these are relative indicators with respect to the demographics of a country. These indicators are irrelevant regarding the nature of the international MOOC market which is characterised by highly scalable business models with significant network effects. These markets often tend to develop towards oligopolistic or even monopolistic structures. In terms of MOOC, the US providers, both platforms as well as content providers, already have a significant head start and will presumably further leverage on it. So, most probably they remain the key players.

Another aspect is how the MOOC market will develop as a whole in future. Presently, the MOOC hype seems to be cooling down (e.g., Kolowich 2015). So, it will be interesting to pursue how the MOOC providers further develop their business models to make them sustainably profitable, whether by subsidies, course fees or other sources of income. Last but not least, long term acceptance of MOOC will be interesting to watch - will MOOC vanish due to unsustainable business models or a decline of interest from students; will they poach from the traditional education industry or will they complement classic educational offers and peacefully coexist with them.